\documentclass[11pt,twoside]{article}
\usepackage{asp2010}

\resetcounters

\begin{document}

\title{Astrophysics Source Code Library: Incite to Cite!}
\author{Kimberly~DuPrie$^1$, Alice~Allen$^1$, Bruce~Berriman$^{2,3}$, Robert~J.~Hanisch$^{3,4}$, Jessica~Mink$^5$, Robert~J.~Nemiroff$^6$, Lior~Shamir$^7$, Keith~Shortridge$^8$, Mark~B.~Taylor$^9$, Peter~Teuben$^1$$^0$ and John~F.~Wallin$^1$$^1$
\affil{$^1$Astrophysics Source Code Library}
\affil{$^2$Infrared Processing and Analysis Center, California Institute of Technology}
\affil{$^3$Virtual Astronomical Observatory}
\affil{$^4$Space Telescope Science Institute}
\affil{$^5$Harvard-Smithsonian Center for Astrophysics}
\affil{$^6$Michigan Technological University}
\affil{$^7$Lawrence Technological University}
\affil{$^8$Australian Astronomical Observatory}
\affil{$^9$University of Bristol}
\affil{$^1$$^0$Astronomy Department, University of Maryland}
\affil{$^1$$^1$Middle Tennessee State University}}

\begin{abstract}

The Astrophysics Source Code Library (ASCL, \url{http://ascl.net/}) is an on-line registry of over 700 source codes that are of interest to astrophysicists, with more being added regularly. The ASCL actively seeks out codes as well as accepting submissions from the code authors, and all entries are citable and indexed by ADS. All codes have been used to generate results published in or submitted to a refereed journal and are available either via a download site or from an identified source. In addition to being the largest directory of scientist-written astrophysics programs available, the ASCL is also an active participant in the reproducible research movement with presentations at various conferences, numerous blog posts and a journal article. This poster provides a description of the ASCL and the changes that we are starting to see in the astrophysics community as a result of the work we are doing.
\end{abstract}

\section{Brief history of the ASCL}
The ASCL was created in 1999 and by 2002 had collected 37 codes after which no new codes were added for several years. In 2010 it was resurrected and moved to Starship Asterisk, the discussion forum for Astronomy Picture of the Day (APOD).\footnote{Starship Asterisk:
\url{http://asterisk.apod.com/}}  Since the move it has added an average of 17 codes a month. In 2011 an Advisory Committee composed of scientists from various institutions was established to provide guidance for the expansion of the revamped ASCL.  Since 2012, ASCL codes have been incorporated into ADS.

\section{Goals of the ASCL}
One of the cornerstones of scientific research is reproducibility. Given the advances in technology, having access to the data alone is no longer enough: scientists need to know how the data were processed in order to reproduce the results. Unless the data processing is very simple, the only feasible way to reproduce the results is to use the same software that the original researcher used. Oftentimes it is not immediately obvious where to obtain this software, or even if the software is available at all. The key goal of the ASCL is to improve the transparency of research by providing a way to link papers to software (and software to papers) and to provide an easy way to find codes.

A secondary goal of the ASCL is to promote code sharing and reuse. If the software you need to solve a problem already exists then why not make use of it? Unfortunately it is often difficult to know if the software exists or where to find it. Entries in the ASCL are fully searchable (code name, description and author) making code discovery easy, and since all software listed is linked to a refereed article users can be confident about the quality of the code. Additionally, ASCL entries are indexed in ADS, providing another avenue for code discovery as well as making the code easily citable.

\section{How we do what we do}
Common problems faced by on-line registries of any kind are out-of-date or static data, broken links and visibility. The ASCL addresses all of these issues:

\begin{itemize}
\item We actively seek out codes as well as soliciting entries from code authors.
\item  We regularly run a link-checker to identify broken links. When broken links are found the editors will search out the new link or ask the code authors to provide an updated URL.
\item Taking advantage of the fact that the ASCL is hosted on the APOD site, we periodically post a link from APOD to ASCL in order to increase our exposure.
\item When the editors add new entries to the ASCL they notify the code authors via email. This not only gives the authors the opportunity to verify their entry, it also serves as a way to increase visibility of the ASCL.
\item We occasionally post articles to sites such as AstroBetter\footnote{AstroBetter: \url{http://www.astrobetter.com/}} and Astronomy Computing Today\footnote{Astronomy Computing Today: \url{http://astrocompute.wordpress.com/}} to inform people about the ASCL and the importance of reproducibility.
\item Representatives of the ASCL regularly attend conferences such as ADASS and the meeting of the American Astronomical Society (AAS) and present posters, give talks, hold discussions and give demos.
\item The ASCL has its own Facebook and Google+ pages where we regularly post updates about the codes that have been added.
\end{itemize}

\section{Update on growth and activity}
Although the number of codes posted by their authors is slowly increasing (7 in 2013 through August compared to 1 for all of 2010) most of the codes are still being added to ASCL by the editors. Figure~\ref{Entries} shows how ASCL is continuing to grow.

\articlefigure{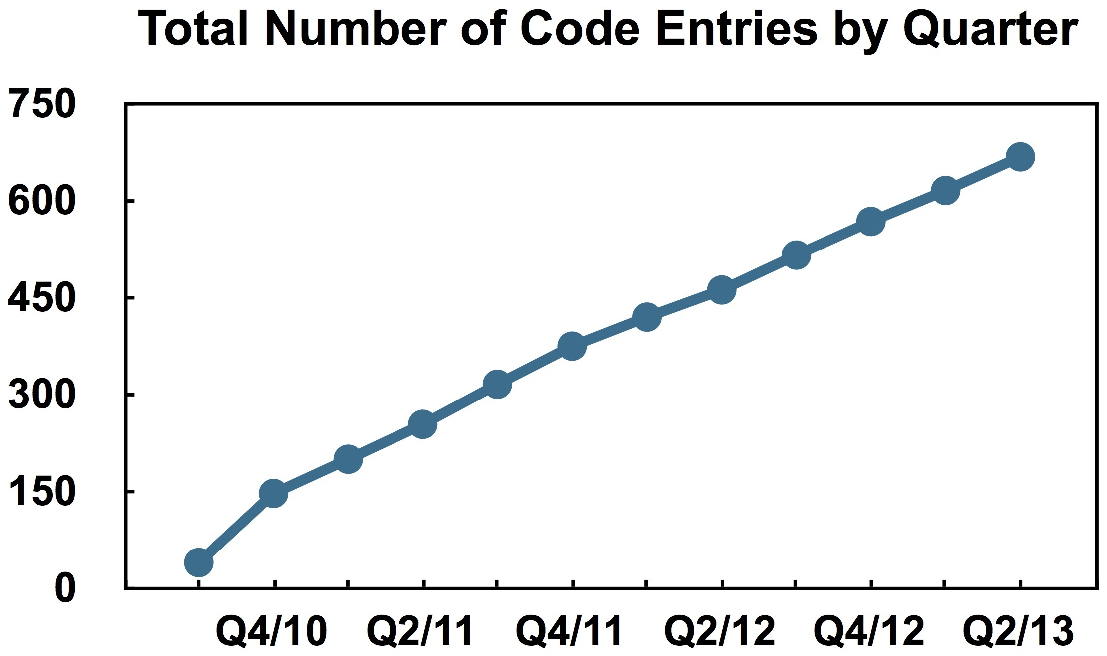}{Entries}
{Total number of code entries by quarter, Q3 2010 - Q2 2013.}

Similarly we are seeing an increase in the number of pageviews, as shown in Figure\ref{Hits}. Although this graph is more spiky, there is still a clear upward trend. Using Google Analytics, our preliminary analysis shows hits from over 60 different countries.

\articlefigure{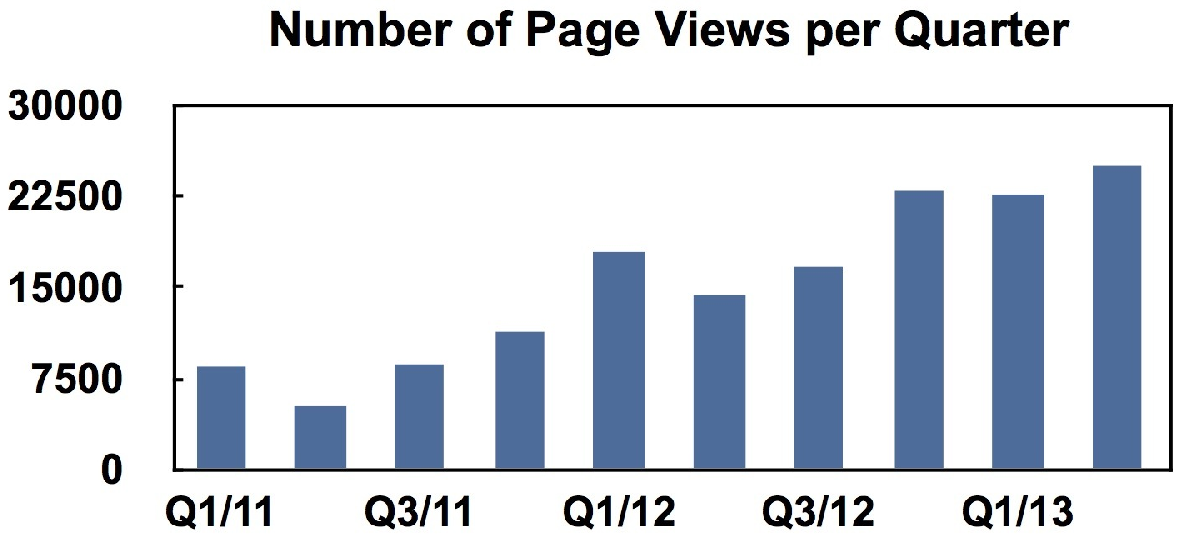}{Hits}
{Pageviews per quarter, Q1 2011 - Q2 2013.}

\section{Changes in the community}
Before the ASCL there was no standard way to cite code in a paper. Some software sites request people to cite a particular paper or to provide a link to their website. Although this ad-hoc citation format allows code authors to track the number of citations for their code, it is difficult to do any analytics on the number of codes that are cited in general, or to easily compare the number of citations for different pieces of code. The standard citation format provided by ASCL addresses both of these issues. Since we started indexing ASCL entries in ADS we have seen a gradual increase in the number of citations: at the time of this writing there are 64 citations for our entries. Although authors could cite software before the ASCL provided this capability, a standard citation mechanism that supports indexing will allow code authors to easily track citations for their code, and therefore we believe will encourage more people to do so. We will need to keep an eye on the statistics to see if this proves to be true.

The ASCL is actively participating in workshops and discussion forums regarding code sharing. We have organized meetings with code authors to identify the obstacles to code sharing as well as possible solutions. The "Bring out your codes!" BoF at ADASS XXII \citep{bof_adassxxii} and the "Astrophysics Code Sharing?"\footnote{Blog post: \url{http://tinyurl.com/code-sharing-splinter}} splinter meeting at AAS 221 saw particularly lively discussions, and followup meetings have been scheduled. In addition to meetings such as these, the ASCL gave a presentation at the Preserving.exe summit at the Library of Congress stressing the importance of code preservation. By making people more aware of the importance of code preservation and sharing, addressing obstacles they may face, and giving them a vehicle to promote their codes we expect to continue to see an increase in activity in this area.  

\bibliography{P069}

\end{document}